\documentstyle[twocolumn,prl,aps]{revtex} 
\begin{document}
\draft

\twocolumn
[\hsize\textwidth\columnwidth\hsize\csname @twocolumnfalse\endcsname

\title{\bf
Excitation Spectrum and Superexchange Pathways in the Spin Dimer VODPO$_4\cdot 1/2 $D$_2$O
}

\author{
D.A.Tennant$^1$,  S.E.Nagler$^1$, A.W.Garrett$^2$, T.Barnes$^3$
and C.C.Torardi$^4$
}

\address{
$^1$Solid State Division,
Oak Ridge National Laboratory, \\
Oak Ridge, TN 37831-6393  \\  
$^2$Department of Physics,
University of Florida,
Gainesville, FL 32611-0448\\
$^3$Theoretical and Computational Physics Section, 
Oak Ridge National Laboratory, \\
Oak Ridge, TN 37831-6373,  \\  
Department of Physics and Astronomy,
University of Tennessee, \\
Knoxville, TN 37996-1501 \\
$^4$DuPont Central Research and Development\\
P.O.Box 80356, Wilmington, DE 19880-0356\\
}

\maketitle

\begin{abstract}
Magnetic excitations have been investigated in the 
spin dimer material VODPO$_4 \cdot 1/2$D$_2$O\
using inelastic neutron scattering. 
A dispersionless magnetic mode was 
observed at an energy of 7.81(4)~meV.
The wavevector dependence of the scattering intensity
from this mode
is consistent with the excitation of isolated 
V$^{4+}$ spin dimers with a V-V
separation of 4.43(7) \AA. 
This result is unexpected since the V-V pair 
previously thought to constitute the
magnetic dimer has a separation of 3.09 \AA.
We identify an alternative V-V pair as the likely magnetic dimer,
which involves superexchange pathways through 
a covalently bonded PO$_4$ group. 
This surprising result casts doubt on the interpretation
of (VO)$_2$P$_2$O$_7$\ as a spin ladder. 
\end{abstract}

]
\vspace{0.3cm}

One of the goals of modern condensed matter physics
is the comprehensive 
understanding of the ground state structure and 
excitations in quantum magnetic systems.
Low dimensional systems, including spin chains,
planes, and intermediate dimensional spin ladders are 
of particular importance because of the role they may
play in the physics of high-T$_{C}$ superconductors \cite{slad}.
The exchange coupled spin dimer is the simplest interacting 
quantum magnetic system. Spin dimers can be thought of
as the basic building blocks for more complicated spin
systems, including RVB spin liquids and spin ladders.
In this Letter we present the results of a neutron scattering
study of excitations in a dimer system of particular
interest: deuterated vanadyl phosphate demihydrate,
{VODPO$_4 \cdot 1/2$D$_2$O\ , (abbreviated VODPO). VODPO is closely 
related to the compound (VO)$_2$P$_2$O$_7$\ (VOPO) \cite{vopo1st,roger}, 
which is widely
thought to be an excellent realisation of the two-leg 
antiferromagnetic Heisenberg spin ladder.
The present results show that there has been a fundamental 
misunderstanding of the magnetism in these materials,
and that a careful re-examination of conclusions
drawn about VOPO is called for. In particular the dominant
exchange couplings have been misidentified.

The spin ladder nature of VOPO was proposed because
the crystal structure contains $S=1/2$ V$^{4+}$ ion 
``rung'' pairs which are stacked into ladders \cite{vopo1st}.
Magnetic interactions were thought to occur through 
V-O-V superexchange paths along both rung and chain 
directions. Neighbouring ladders are separated by
PO$_{4}$ complexes which have been assumed to be
magnetically inert. (VO)$_2$P$_2$O$_7$\ can be synthesized by
dehydration of the precursor material VOHPO$_4 \cdot 1/2$H$_2$O\ (VOHPO) 
\cite{tc,jjjb}. VOHPO contains V-V pairs identical to the proposed
rungs of the VOPO ladder, but magnetically decoupled 
along the chain direction (see Fig. 1). It is 
therefore possible to study part of the ladder 
in isolation in VOHPO or the deuterated counterpart
VODPO. For neutron scattering experiments the 
deuterated material has the advantage of lowering the 
incoherent background without significantly
altering the magnetic interactions.

An idealized isotropic spin-dimer material consists of isolated pairs of
magnetic ions that interact through a Heisenberg Hamiltonian,

\begin{equation}
H = J \, \vec S_1 \cdot \vec S_2  \  .
\end{equation}
Johnson {\it et al.} \cite{jjjb} measured the magnetic susceptibility of
VOHPO and found that it is accurately described by the Bleaney-Bowers
formula for the susceptibility of isolated dimers of $S=1/2$ spins,
with 
$J=7.6$~meV
and $g~=~1.99$ (close to the free electron value, as expected for
an orbitally quenched V$^{4+}$ ion). 

By measuring the $E$ and $Q$ dependence of magnetic excitations
using inelastic neutron scattering we can test this
simple Hamiltonian and deduce the distance between V-V ion pairs in a dimer.
Since a dimer has an $S_{total}=0$ ground state
and only one excited level (an $S_{total}=1$ triplet at 
$E_1-E_0 = J$), 
inelastic neutron scattering should reveal a single, dispersionless
excitation. The susceptibility fit predicts that this
excitation should appear at $E_{gap}=7.6$~meV. 

On carrying out a powder average over the neutron scattering structure
factor $S(\vec Q,E)$ of a dimer, the powder scattering intensity $I(Q,E)$
is found to be \cite{fu}

\begin{equation}
I(Q,E) \propto |F(Q)|^2 \bigg( 1 - {\sin{QR} \over QR }\bigg) \; \delta(E-E_{gap})   \ .
\end{equation}
Although we expect only a single excited level, the 
$Q$ dependence of the magnetic excitation is especially interesting
in that it allows a determination of the separation $R$ 
of the magnetic ions forming the dimer pair.

The VODPO used in this study was synthesised from
V$_2$O$_5$, D$_3$PO$_4$, and D$_2$O, which were
refluxed under nitrogen for 2 days to obtain
VOPO$_4$$\cdot$2D$_2$O. This product was heated
in a Teflon-lined reaction vessel, with isopropanol-d8
under nitrogen and autogenous pressure at 125$^{\circ}$
for 7 days, and then was filtered and dried in a nitrogen glove bag to
obtain VODPO. Infrared spectroscopy indicated an
H/D ratio of 0.04. 

Neutron powder diffraction
was used to refine the crystal
structure, which gave parameters similar to previous results
for VOHPO \cite{tc}.
Rietveld refinement yielded an orthorhombic structure at T = 10 K,
with lattice parameters of
$a=7.4102(6)$ \AA, $b=9.5861(8)$ \AA, and $c=5.6873(7)$~\AA.
It also confirmed the atomic fraction of
deuterium as being at least 96\%.
The crystal structure of VODPO at 10 K is illustrated in Figure~1a.
The structure contains V$^{4+}$ ions,
each of which has a single $3d$ electron
and effective spin $S=1/2$. Neighboring V$^{4+}$ ions
belong to edge-sharing VO$_{6}$ octahedra.
Next-nearest-neighbor V$^{4+}$ ions are separated
by PO$_{4}$ tetrahedra.

Possible exchange paths
between neighboring V$^{4+}$ ions are shown
in Figure~1b.
The nearest-neighbor pathway (P1) has a V-V separation of
$d_{\rm VV}=3.09$ {\AA}
and the next-nearest-neighbor pathway (P2) has a
V-V separation of $d_{\rm VV}=4.32$ {\AA}. Consideration
of orbital overlaps led Beltr\'an-Porter {\it et al.}
\cite{euro_chemists} to conclude that superexchange
in VODPO is probably dominated by the P1 pathway.

The magnetic excitation spectrum of VODPO was measured using the HB1
triple-axis spectrometer at the High Flux Isotope Reactor at
Oak Ridge National Laboratory.
Approximately 10 grams of VODPO powder was sealed in an aluminum cylinder
and mounted in a displex cryostat, allowing
the temperature to be varied over
the range 10 K $-$ 300 K, with an accuracy of $\pm$0.1~K. 
The $Q$ resolution of the spectrometer was relaxed 
because the excitations were expected to be dispersionless.
A vertically focussed
PG(002) monochromator and horizontally focussed PG(002) analyser were used,
and an instrumental collimation of $48'$-$40'$-open-open was chosen.
Measurements were made using a fixed final energy $E_F=13.5$~meV, giving an 
energy resolution of 1.16 meV full-width half maximum (FWHM).

Scans in $E$ at fixed $Q$ were carried out 
in order to measure the magnetic excitations. 
Figure~2 shows the scattering at temperatures of 
10 K, 50 K and 200 K, at a constant wavevector transfer of $Q=1.0$ \AA$^{-1}$.
In each scan there is a well-defined peak near
7.8 meV, close to the value of 7.6 meV anticipated
from the susceptibility
data \cite{jjjb}.
To extract accurate positions and widths for the peaks,
a least-squares routine
was used to fit a Gaussian profile with a
linearly-sloping background.
This gave a peak center of
$J=7.86(6)$ meV and FWHM of 2.05(10)~meV for the
scan at 10 K.
Peak centers, intensities and widths were extracted at
the other temperatures using the same procedure.  
The integrated intensity falls with increasing 
temperature roughly as expected for a magnetic dimer mode.

To investigate the dispersion of this mode we
carried out a series of scans at 10 K with
increasing wavevector,
covering the range $Q=1.0$-$4.5$ \AA$^{-1}$, at
intervals of $\Delta Q=0.1$ \AA$^{-1}$.
Each scan was fitted to a Gaussian
with linear background.
The mode was found to be essentially non-dispersive,
although the fitted
width increases slightly with $Q$.

To find the dimer spacing $d_{\rm VV}$,
a constant energy scan was made at 
7.75 meV. Figure~3
shows this scan at a temperature of 10 K.
There are two clear  maxima in the data,
near 1.0 \AA$^{-1}$ and $2.3$ \AA$^{-1}$;
this oscillatory behavior is expected for scattering
from a dimer, {\it cf.} Eq. (2).
The solid line in the
figure is a least-squares fit to Eq. (2), 
in which the
${\langle}J_{0}{\rangle}$ approximation for the $ V^{4+} $
form factor \cite{brown} has been used.
A sloping background
has been added to take account of the phonon
contribution, which increases with $Q$.
The fit yields a V-V separation
in the dimer of $ d_{\rm VV}=4.43(7)$ \AA, which we consider 
our definitive value.

Figure~4 shows a two-dimensional color plot of the combined constant-$Q$ scans
at 10 K.
The data have been interpolated to smooth the image, and the
intensity is indicated by a colorbar.
The spin excitation is observed as a
band at 7.8 meV, with maximum intensity at the lower wavevectors.
Two maxima of the structure factor as a function of $Q$ 
appear as bright regions near
$1.0$~\AA$^{-1}$ and $2.3$~\AA$^{-1}$.
The non-magnetic phonon background is visible 
at large $Q$, especially  
at energies above 8 meV.
A simultaneous least-squares fit was made to
Eq. (2) plus a background term linear in $E$ 
and $Q$.  The delta function in
energy
was replaced by a Gaussian lineshape.
The global fit gave an excellent description of the data
with $J=7.81(4)$ meV, $d_{\rm VV}=4.49(8)$ \AA, consistent
with the $4.43(7)$ \AA found in the constant-$E$ scan in Figure~3.
A goodness-of-fit reduced $\chi^{2}=1.12$.
The fitted model is shown as a contour plot (solid lines)
overlying the data in Figure~4.

Our constant-$E$ measurement gave a dimer spacing of $d_{\rm VV}=4.43(7)$ \AA,
consistent with the exchange path P2
shown in Figure~1b ($d_{\rm VV}=4.32$ {\AA}), but not with
the shorter P1 exchange path ($d_{\rm VV}=3.09$ {\AA}) 
assumed to be dominant in earlier work \cite{jjjb,euro_chemists}.
This is surprising because it implies that the exchange is stronger through
a V-O-P-O-V pathway than through a much shorter V-O-V pathway. 
The conventional understanding of superexchange is that
the interaction decreases rapidly with the number of ions in
the exchange path \cite{goodenough}.
We argue however that when there is a
covalent complex present, such as the PO$_{4}$ in VODPO,
coherent orbitals
can provide a strong exchange route. Understanding
the electron wavefunctions within such covalent groups
is therefore crucial in predicting the important superexchange pathways.
Presumably the 
V-O-V superexchange is also weakened by the large departure of the bond angle
(experimentally 97.5$^\circ$)
from the optimum 180$^\circ$ for -O-.

Beltr\'an-Porter {\it et al.} \cite{euro_chemists} have
considered exchange pathways in several vanadyl phosphates.
Although they did not anticipate 
a dominant PO$_4$ pathway in VODPO,
they do argue that PO$_{4}$ pathways may be significant 
in (VO)$_2$P$_2$O$_7$, as well as V-O-V.
If V-O-V (rung) and V-O-P-O-V are
the important exchange pathways
in (VO)$_2$P$_2$O$_7$,
its magnetic Hamiltonian
may be better approximated by
an alternating chain than a ladder,
with the alternating chain directed along the rung of the ladder
({\it cf.} Johnston {\it et al.} \cite{vopo1st}).
Our results on VODPO confirm the importance of PO$_4$ pathways,
and our recent study 
of magnetic excitations in 
(VO)$_{2}$P$_{2}$O$_{7}$ single crystals \cite{vopornl} appears to 
support 
an alternating-chain model.

In summary, we have used inelastic neutron scattering 
on a powder sample to study
magnetic excitations in VODPO$_4 \cdot 1/2$D$_2$O.
Our results support a magnetic dimer model for this system,
with an exchange constant of $J=7.81(4)$~meV, close to the
value found in a susceptibility fit. The V-V spacing
within the dimer however is found to be 4.43(7)~\AA, which indicates
that the dimer pair has been misidentified in previous work.
The dimer pair we identify has a V-O-P-O-V superexchange pathway 
through a 
PO$_4$ group, which evidently leads to 
a stronger magnetic interaction than
the shorter V-O-V pathway.

Strong superexchange through covalently bonded
inorganic complexes 
is presumably important in many magnetic insulators. For example,
an exchange of $\approx 30$ meV 
through
WO$_{6}$ complexes 
has recently been identified 
in CuWO$_4$ \cite{lake} and -O-Mo-O- pathways have been
noted in FeMoO$_4$Cl \cite{tclr}.
The various topologies of covalent groups should allow
the synthesis of a wide range of low-dimensional
magnetic materials \cite{NATO}.

\newpage
\acknowledgements

We would like to acknowledge useful conversations with
Joe Cable and Bryan Chakoumakos.
We thank Pengcheng Dai and Jaime Fernandez-Baca for
valuable assistance with HB1.
Expert technical assistance was provided by Scott Moore and Brent Taylor.

This work was supported in part by the United States Department
of Energy under contract DE-FG05-96ER45280 at the University of Florida, and by
Oak Ridge National Laboratory, managed for the U.S. D.O.E.
by Lockheed Martin Energy Research Corporation
under contract DE-AC05-96OR22464.

\newpage 

\begin{center}
{\Large Figure Captions}
\end{center}

\begin{figure}
{Fig.1. (color)
(a) The structure of VODPO at 10K is shown. Edge-sharing
VO$_{6}$ octahedra (green) are connected by
PO$_{4}$ tetrahedra (red). These layers are separated
by D atoms (grey). An orthorhombic unit cell
is indicated by the dashed line, and the
crystal axes are indicated at lower left. (b) This shows a schematic
projection of the
crystal structure in the $a-b$ plane illustrating possible
superexchange pathways between V$^{4+}$ ions. V ions are
shown in green, P in red, and O in blue.
P1 is a V-O-V pathway (d$_{\rm VV}$=3.09 \AA),
and P2 is V-O-P-O-V (d$_{\rm VV}$=4.32 \AA).
}
\end{figure}

\begin{figure}
{Fig.2.
Scans in energy at a constant wavevector transfer of $Q=1.0$ {\AA}$^{-1}$
are displayed.
The data are scaled to a fixed number of incident neutrons, as
determined by a monitor counter. Monitor 100 corresponds to roughly
1 minute of counting time per point. The error bars represent one
standard deviation.  The profiles at different temperatures
have been offset by a constant number of counts for clarity of presentation.
The solid lines are fits to a Gaussian plus a sloping background as
discussed in the text.
}
\end{figure}

\begin{figure}
{Fig.3.
A scan in wavevector transfer $Q$ at a constant energy transfer of
7.75 meV (essentially the dimer excitation energy) is shown, with some 
representative error bars.
The solid line is a fit to the dimer model,
as described in the text.
}
\end{figure}

\begin{figure}
{Fig.4. (color)
This shows a two-dimensional color plot of the combined constant-$Q$ scans.
The contour lines are the results of the global
fit described in the text. Counts per monitor 100 are given by
the colorbar.
}
\end{figure}
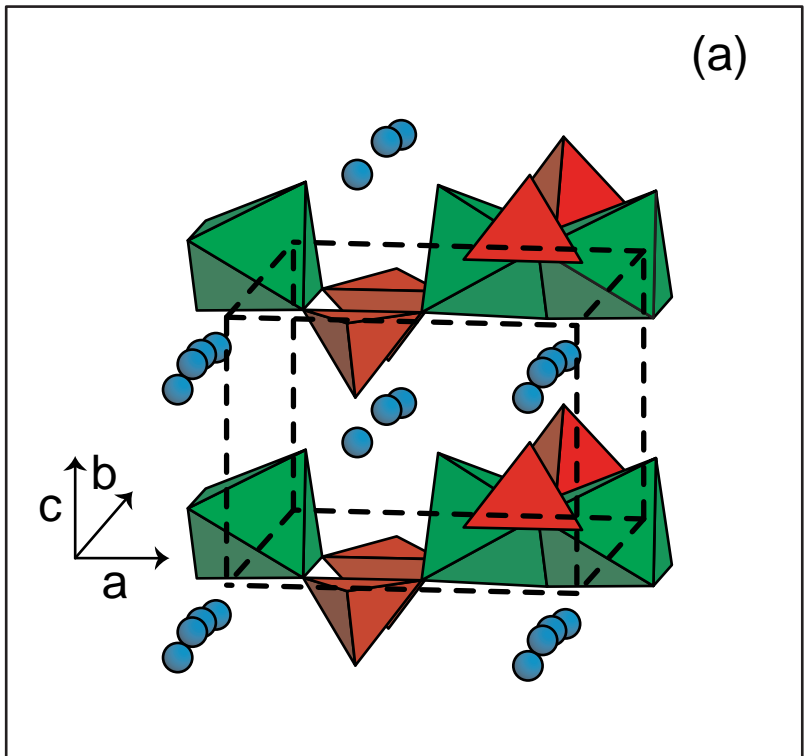
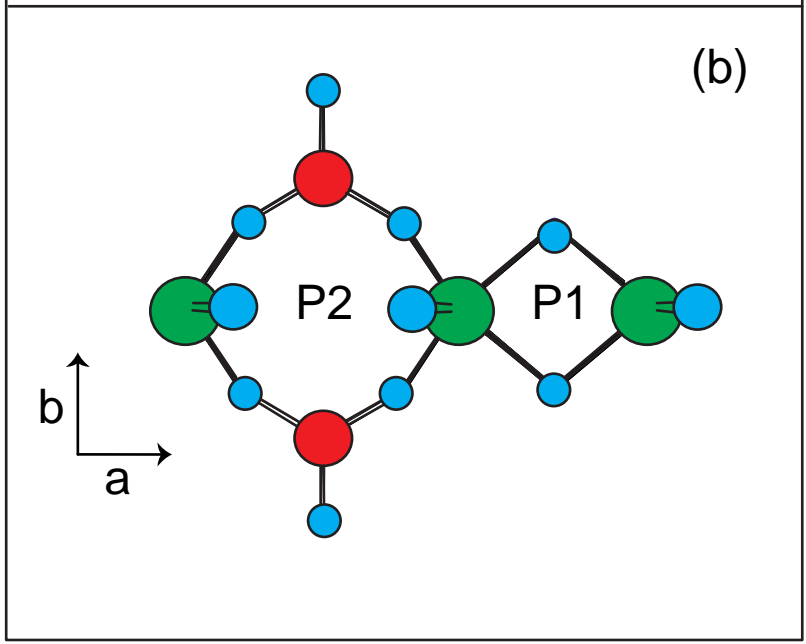
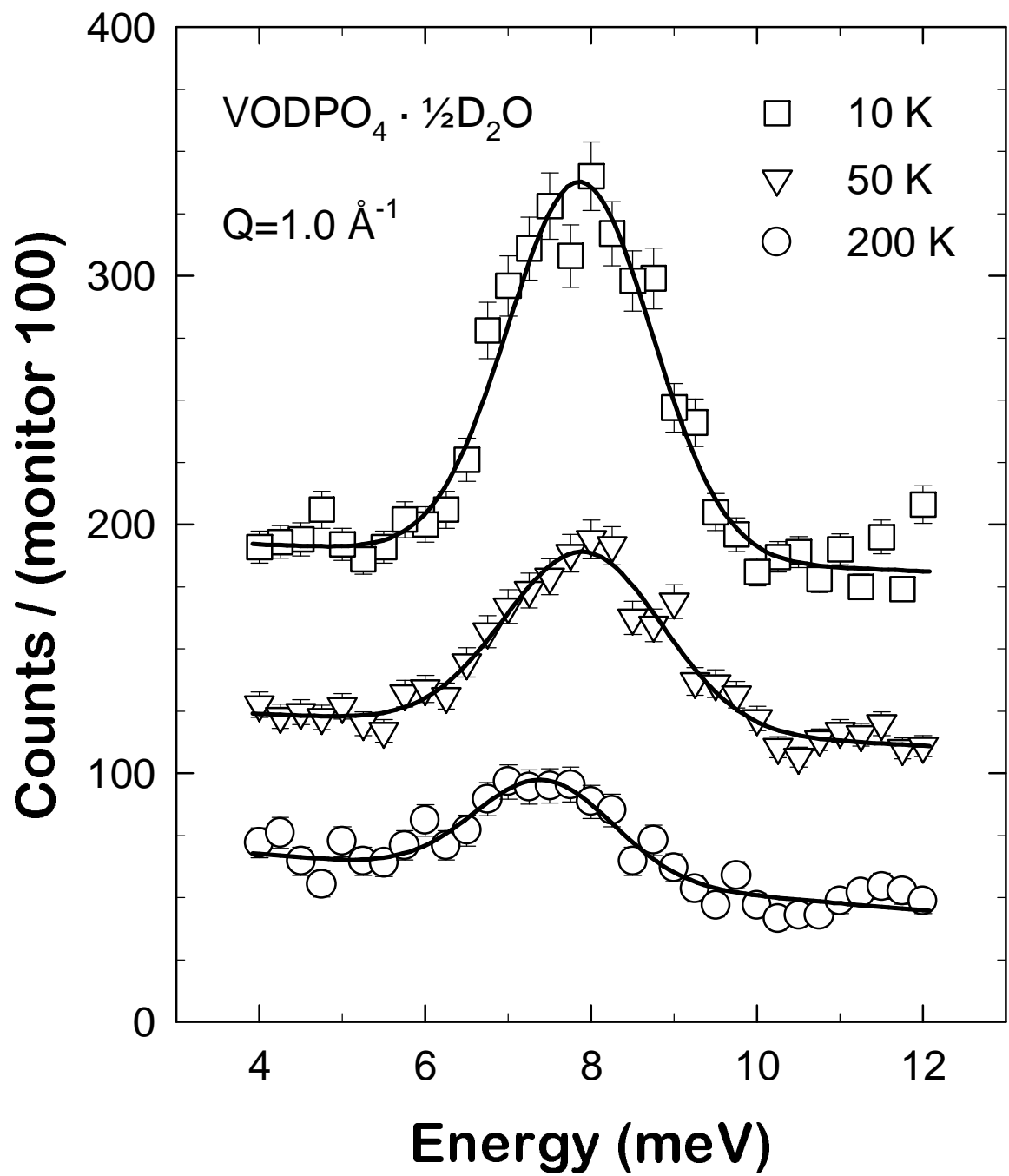
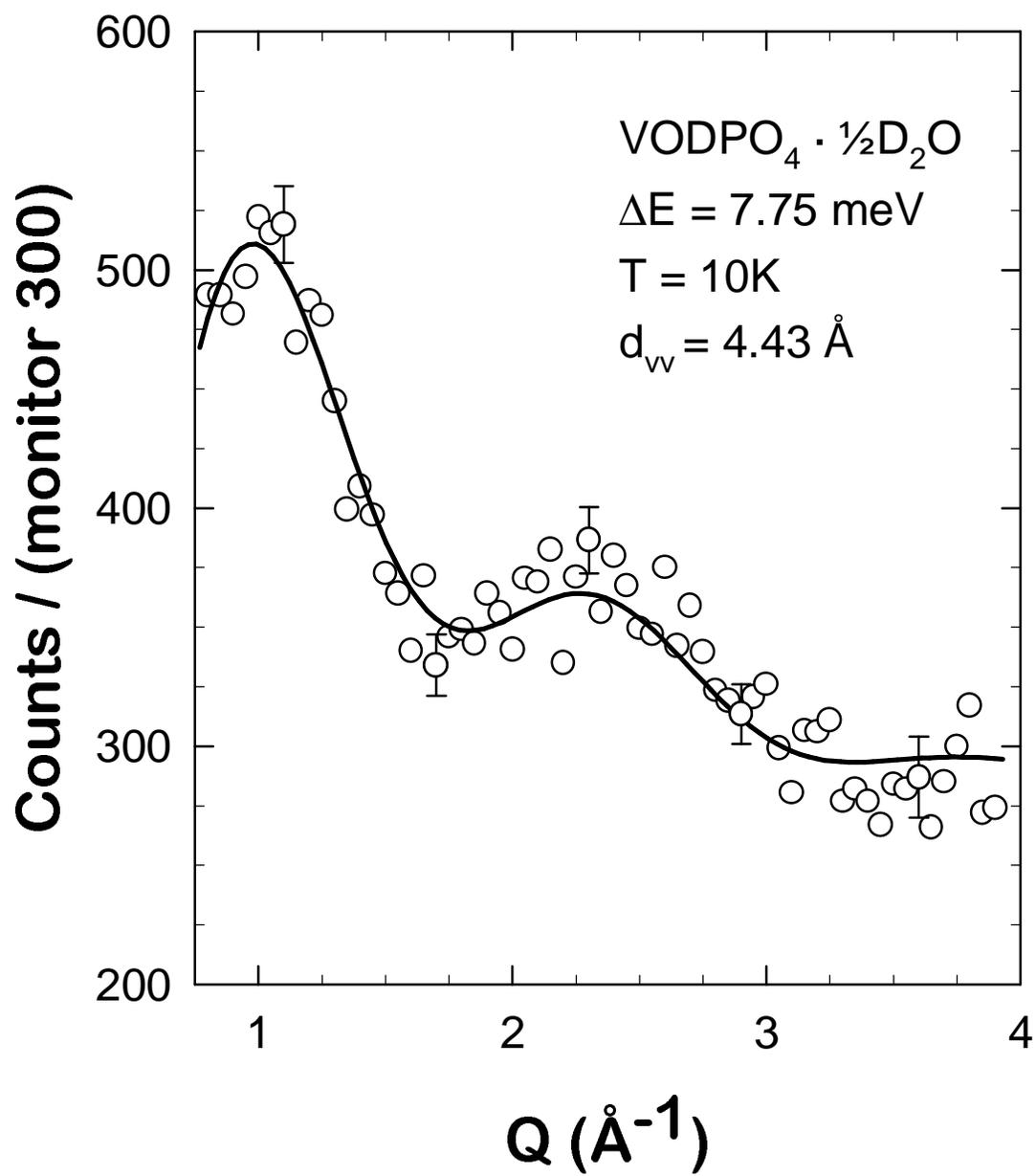
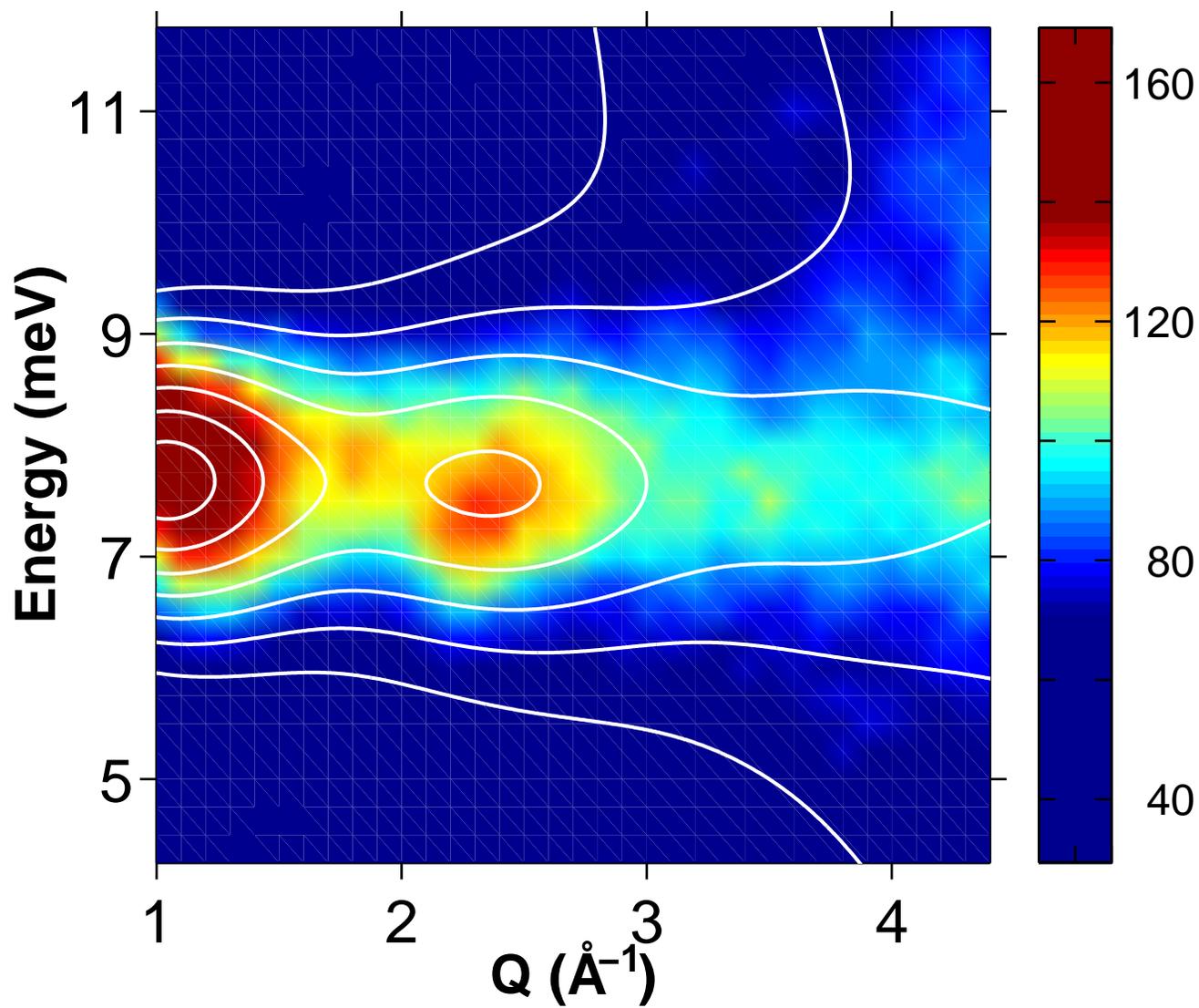
\end{document}